\documentclass[prl,reprint]{revtex4-2}

\usepackage[dvipsnames]{xcolor}
\usepackage[colorlinks,linktocpage]{hyperref}
\hypersetup{
	citecolor  = NavyBlue,
	linkcolor  = NavyBlue,
	urlcolor = NavyBlue
}

\usepackage{amsmath,amssymb}
\usepackage{graphicx}
\usepackage{enumitem}

\newcommand{\hl}[1]{\textcolor{BrickRed}{#1}}

\def\newpar{\vskip4pt}

\makeatletter
\def\frontmatter@above@affiliation{\addvspace{3pt}}
\makeatother

\newcounter{appendix}
\renewcommand{\theappendix}{\Alph{appendix}}
\newcommand{\appsection}[1]{
	\refstepcounter{appendix}
    \renewcommand{\theequation}{\theappendix\arabic{equation}}
    \setcounter{equation}{0}
    \paragraph{Appendix \theappendix: #1}---\,
  }

\begin{document}

\title{Optimal Architecture and Fundamental Bounds\\ in Neural Network Field Theory}

\author{Zhengkang Zhang}
\affiliation{Department of Physics and Astronomy, University of Utah, Salt Lake City, UT 84112, USA}

\begin{abstract}
Neural network field theory (NNFT) represents fields as neural networks and samples field configurations by drawing network parameters from a probability distribution.
We identify a previously unexplored architectural freedom in NNFT, parameterized by $\alpha$, that leaves the infinite-width theory invariant but dramatically affects finite-width errors in the calculation of correlation functions.
For a massive scalar field, we show that $\alpha=0$, corresponding to propagator-weighted neuron momenta and constant neuron amplitudes, is optimal: it minimizes finite-width variance and uniquely removes IR-sensitive corrections in the interacting theory.
Even at $\alpha=0$, relative errors from both bias and variance grow exponentially with distance beyond the correlation length.
The bias can be removed by extrapolating to infinite width, which we demonstrate numerically, while the variance imposes a fundamental bound on the achievable signal-to-noise ratio as in lattice field theory.
These results chart a path toward developing NNFT into a practical tool for the numerical study of field theories.
\end{abstract}

\maketitle

\paragraph{Introduction}---\,
Artificial neural networks and Euclidean field theories can both be viewed as collections of random variables.
A neural network computes a function of its input, and when the network parameters are drawn from a probability distribution, the output is a random function of the input.
A field theory, meanwhile, is a prescription for computing correlation functions of a random field, i.e., a random function of spatial coordinates, $\phi(x)$, where the probability distribution over field configurations is determined by the action $S[\phi]$.
Given this structural similarity, a natural question arises: can neural networks with random parameters be used to simulate field theories?

To answer this question, we need to understand how, given a field theory action, one can construct a neural network architecture and a probability distribution over its parameters such that the resulting random function reproduces the statistics of the field theory, i.e., its correlation functions.
This is the core objective of the emerging neural network field theory (NNFT) program~\cite{Halverson:2020trp,Halverson:2021aot,Demirtas:2023fir}.
Since its inception, NNFT has been extended in many directions, including conformal field theories~\cite{Halverson:2024axc,Capuozzo:2025ozt,Robinson:2025ybg}, fermions and supersymmetry~\cite{Huang:2025ipy,Frank:2025zuk}, and string theory~\cite{Frank:2026bui,Ageev:2026ofv}; see also Refs.~\cite{Erbin:2021kqf,Erbin:2022lls,Howard:2024kfd,Halverson:2024hax,Ferko:2025ogz,Ferko:2026axm,Ageev:2026qyh,Ferko:2026ken}.
In contrast to traditional lattice field theory which discretizes spacetime, NNFT represents the field as a neural network and samples field configurations by drawing network parameters.
This offers potential advantages: the fields live directly in the continuum, and exact Euclidean symmetry can be preserved~\cite{Maiti:2021fpy}.

The starting point of NNFT is the observation that, by the central limit theorem, a neural network with infinite width (i.e., infinitely many neurons per layer) and independently drawn parameters produces a Gaussian random field, i.e., a free field.
This is a field-theoretic restatement of the foundational result in machine learning (ML) that infinitely wide neural networks are Gaussian processes~\cite{Neal1996,Williams1996,Lee:2017qzq,Matthews2018,Yang2019,Hanin2021}.
To obtain interacting theories, one can break the statistical independence of network parameters while staying in the infinite-width limit~\cite{Demirtas:2023fir}.

In practice, however, the network width $N$ is finite, which introduces a systematic bias in the form of $1/N$ corrections to the correlation functions.
Moreover, the finite ensemble size of Monte Carlo sampling gives rise to statistical variance.
A precise characterization of these errors and the limitations they impose is essential for developing NNFT into a practical computational tool.

In this letter, we identify a previously unexplored freedom in NNFT architectures that has dramatic consequences for these errors.
We focus on a prototypical massive scalar field theory and parameterize this freedom by $\alpha$, which corresponds to different ways of distributing the momentum-space integrand between the probability distribution over network parameters and the neuron amplitude.
All choices of $\alpha$ yield the same field theory in the $N\to\infty$ limit, but they give qualitatively different predictions at finite $N$.
Previous work~\cite{Halverson:2021aot,Demirtas:2023fir} adopted $\alpha=-1$; this particular choice gives rise to $(\Lambda/m)$-enhanced bias for correlation functions at distances $r\sim 1/m$~\cite{Sen:2025vzl}, where $\Lambda$ is the UV cutoff and $m$ is the mass of the scalar field.
Here, we systematically extend these results by considering both bias and variance, for all $\alpha$ at all $r$.

We find that, at $r\gtrsim 1/m$, both bias and variance grow exponentially with $mr$ relative to the signal.
This exponential growth represents a qualitatively more severe obstacle than the polynomial enhancement previously identified at $r\sim 1/m$.
The bias is a systematic error that can be removed by extrapolating to $N\to\infty$.
The variance, by contrast, represents an irreducible noise floor that persists even as $N\to\infty$.
Nevertheless, we show that choosing $\alpha=0$ minimizes finite-$N$ contributions above the noise floor.
We perform numerical experiments to demonstrate both the removal of bias via extrapolation and $\alpha$-dependence of the variance for the free theory.
We also consider the interacting $\phi^4$ theory in the perturbative regime and show that $\alpha=0$ removes IR-sensitive corrections.
Together, these results establish $\alpha=0$ as the optimal architecture for NNFT.


\newpar
\paragraph{A family of architectures}---\,
Consider a one-hidden-layer neural network with cosine activation functions, known as Cos-Net~\cite{Halverson:2021aot,Demirtas:2023fir} (or random Fourier features in ML literature~\cite{Rahimi2007}):
\begin{equation}
\phi(x;\theta) = \frac{1}{\sqrt{N}} \sum_{i=1}^N w_i^{(2)} \cos\bigl(w_i^{(1)}\cdot x + b_i^{(1)}\bigr)\,,
\label{eq:cosnet}
\end{equation}
where $x\in\mathbb{R}^d$ is the input and $\theta = \{w_i^{(1)}, b_i^{(1)}, w_i^{(2)}\}_{i=1}^N$ are the network parameters.
When the first-layer weights $w_i^{(1)} \in \mathbb{R}^d$ are drawn from a rotation-invariant distribution and the biases $b_i^{(1)}$ are drawn uniformly from $[-\pi, \pi]$, correlation functions of the network output are Euclidean invariant.
We assume the second-layer weights $w_i^{(2)}$ are drawn from a distribution that is independent of $b_i^{(1)}$ but may depend on $w_i^{(1)}$.
Defining $a_i \equiv \tfrac{1}{2} w_i^{(2)} e^{ib_i^{(1)}}$ and $k_i \equiv w_i^{(1)}$, we can rewrite Eq.~\eqref{eq:cosnet} as a sum of plane waves with random momenta:
\begin{equation}
\phi(x;\theta) = \frac{1}{\sqrt{N}} \sum_{i=1}^N \bigl(a_i\, e^{ik_i\cdot x} + \mathrm{c.c.}\bigr)\,.
\label{eq:fourier}
\end{equation}

By the central limit theorem, $\phi(x;\theta)$ becomes a Gaussian random field as $N\to\infty$.
The $2n$-point connected correlators scale as $1/N^{n-1}$, provided the parameters $k_i, a_i$ associated with each neuron are drawn independently from the same distribution.
Interactions can be introduced by breaking the independence of network parameters~\cite{Demirtas:2023fir}, $P(\theta) \propto P_0(\theta)\, e^{-S_\mathrm{int}[\phi(\theta)]}$, or equivalently by reweighting the free-theory ensemble,
\begin{equation}
\bigl\langle \phi(x_1) \cdots \phi(x_{2n})\bigr\rangle = \frac{\langle \phi(x_1) \cdots \phi(x_{2n})\, e^{-S_\mathrm{int}} \rangle_0^{}}{\langle e^{-S_\mathrm{int}} \rangle_0^{}}\,,
\label{eq:reweight}
\end{equation}
where $\langle\,\cdots\rangle_0^{}$ denotes the expectation value with respect to the free-theory distribution $P_0(\theta)$, and we have suppressed the dependence on $\theta$ for brevity.
In what follows, we first analyze the bias and variance for correlation functions in the free theory, and then discuss how these results extend to the interacting theory via Eq.~\eqref{eq:reweight}.

With independently and identically distributed (i.i.d.)\ parameters $k_i, a_i$, the two-point function $G^{(2)}(x_1,x_2) \equiv \langle \phi(x_1) \phi(x_2) \rangle_0^{}$ receives identical contributions from each neuron, yielding
\begin{equation}
G^{(2)}(x_1,x_2) = 2\int d^dk_i\, p(k_i)\,\bigl\langle|a_i|^2\bigr\rangle\, e^{ik_i\cdot(x_1-x_2)}\,.
\label{eq:G2}
\end{equation}
Here $p(k_i)$ is the probability distribution over $k_i = w_i^{(1)}$; the average over $b_i^{(1)}$ eliminates the $a_i^2$ and $a_i^*{}^2$ terms, and $\langle|a_i|^2\rangle$ denotes the average of $|a_i|^2 = (w_i^{(2)})^2/4$ over $w_i^{(2)}$ at fixed $w_i^{(1)}$.
No summation over repeated indices is implied throughout this letter.
We can choose $p(k_i)$ and $\langle|a_i|^2\rangle$ to reproduce the propagator in any (reflection-positive) Euclidean-invariant field theory.
For a free scalar with mass $m$, we need
\begin{equation}
	p(k_i)\,\langle|a_i|^2\rangle = \frac{1}{2(2\pi)^d}\frac{f_\Lambda(k_i^2)}{k_i^2+m^2}\,,
\label{eq:matching}
\end{equation}
where $f_\Lambda(k_i^2)$ is a UV regulator that satisfies $f_\Lambda(0)=1,\; f_\Lambda(\infty)=0$\,; e.g., one can take $f_\Lambda(k_i^2) = \Theta(\Lambda^2-k_i^2)$ or $f_\Lambda(k_i^2) = e^{-k_i^2/2\Lambda^2}$.

Eq.~\eqref{eq:matching} does not uniquely determine $p(k_i)$ and $\langle|a_i|^2\rangle$, but only their product~\footnote{This freedom was noted in passing in Ref.~\cite{Frank:2026bui} in the context of the 2d free boson.}.
We consider a family of choices parameterized by $\alpha$:
\begin{align}
p(k_i) &= \frac{1}{(2\pi)^d\,\Omega_\alpha} \frac{f_\Lambda(k_i^2)}{(k_i^2+m^2)^{\alpha+1}}\,,\\[4pt]
\bigl\langle|a_i|^2\bigr\rangle &= \frac{\Omega_\alpha}{2}\,(k_i^2+m^2)^\alpha\,,
\end{align}
where $\Omega_\alpha$ is a normalization factor defined as
\begin{equation}
\Omega_\alpha \equiv \int \frac{d^dk}{(2\pi)^d}\frac{f_\Lambda(k^2)}{(k^2+m^2)^{\alpha+1}}\,.
\label{eq:Omega}
\end{equation}
All choices of $\alpha$ yield the same two-point function, and hence the same $N\to\infty$ theory, but they give different predictions for higher-point correlation functions at finite $N$ as we will see shortly.


\newpar
\paragraph{Finite-width bias}---\,
Turning to higher-point correlation functions, we define $\phi_i(x) \equiv a_i e^{ik_i\cdot x} + \mathrm{c.c.}$, and denote the $2n$-point functions of $\phi$ and $\phi_i$ as $G^{(2n)}$ and $G_i^{(2n)}$, respectively.
For i.i.d.\ neurons, we can express $G^{(2n)}$ in terms of $G_i^{(2n')}$ with $n'\le n$; for example,
\begin{widetext}
\vspace{-4pt}
\begin{align}
G^{(4)}(x_1, \dots, x_4) &= \biggl(1-\frac{1}{N}\biggr) \Bigl(G^{(2)}_i(x_1, x_2) \,G^{(2)}_i(x_3, x_4) + \,\cdots \Bigr)_\text{3 terms} +\frac{1}{N} \, G^{(4)}_i(x_1, \dots, x_4)\,, \label{eq:G4} \\[4pt]
G^{(6)}(x_1, \dots, x_6) &= \biggl(1-\frac{1}{N}\biggr)\biggl(1-\frac{2}{N}\biggr) \Bigl(G^{(2)}_i(x_1, x_2) \,G^{(2)}_i(x_3, x_4) \,G^{(2)}_i(x_5, x_6) + \,\cdots \Bigr)_\text{15 terms} \nonumber\\
&\qquad +\frac{1}{N}\biggl(1-\frac{1}{N}\biggr) \Bigl(G^{(4)}_i(x_1, \dots, x_4) \,G^{(2)}_i(x_5, x_6) + \,\cdots \Bigr)_\text{15 terms} +\frac{1}{N^2} \, G^{(6)}_i(x_1, \dots, x_6)\,, \label{eq:G6}
\end{align}
\end{widetext}
where ``$\cdots$'' denotes all distinct permutations of the coordinates; see Appendix~\ref{app:correlation-functions} for details.
In each equation, the first term is the free-theory prediction rescaled by a factor of $1+\mathcal{O}(1/N)$, while the remaining terms constitute additional contributions to finite-$N$ bias.
The bias from these additional contributions relative to the free-theory prediction is, schematically, controlled by the ratios
\begin{equation}
	\frac{1}{N^{n-1}} \,\frac{G_i^{(2n)}}{\bigl[G_i^{(2)}\bigr]^n} \;\equiv\; \kappa_n\,,
	\label{eq:kappa}
\end{equation}
with $n\ge 2$.
To determine these ratios, we note that
\begin{widetext}
\begin{equation}
G_i^{(2n)}(x_1, \dots, x_{2n}) = \frac12\, \biggl(\frac{\Omega_\alpha}{2}\biggr)^{\!n-1}\!\!\int \frac{d^dk_i}{(2\pi)^d} \,\frac{\bigl\langle|a_i|^{2n}\bigr\rangle}{\bigl\langle|a_i|^2\bigr\rangle^n} \frac{f_\Lambda(k_i^2)}{k_i^2+m^2}\,(k_i^2\!+\!m^2)^{\alpha(n-1)} \sum_\mathrm{P} e^{ik_i\cdot(x_\mathrm{in}-x_\mathrm{out})}\,,
\label{eq:Gi2n}
\end{equation}
\end{widetext}
where the sum runs over all $\binom{2n}{n}$ balanced partitions P of the $2n$ coordinates into ``in'' and ``out'' sets, with $x_\mathrm{in}\equiv\sum_{j\in\mathrm{in}}x_j,\; x_\mathrm{out}\equiv\sum_{j\in\mathrm{out}}x_j$.
In what follows, we take $|a_i| = \sqrt{\Omega_\alpha/2}\,(k_i^2+m^2)^{\alpha/2}$ to be deterministic, so that $\langle|a_i|^{2n}\rangle = \langle|a_i|^2\rangle^n$~\footnote{Sampling $|a_i|$ from a Gaussian as in Refs.~\cite{Halverson:2021aot,Demirtas:2023fir} instead gives $\langle|a_i|^{2n}\rangle = (2n-1)!!\,\langle|a_i|^2\rangle^n$, which amplifies the bias and variance by numerical prefactors but does not change the parametric dependence on $\alpha$.}.

We consider three cases in turn:
\begin{itemize}[leftmargin=20pt]
	\item For $\alpha = 0$, Eq.~\eqref{eq:Gi2n} reduces to a sum of propagators, $G_i^{(2n)} \propto \sum_\mathrm{P} G^{(2)}(x_\mathrm{in}\!-\!x_\mathrm{out})$, each falling off at large distances as $e^{-m|x_\mathrm{in}-x_\mathrm{out}|}$.
	By the triangle inequality, $|x_\mathrm{in}-x_\mathrm{out}| \le |x_{j_1}-x_{j_2}|+\cdots+|x_{j_{2n-1}}-x_{j_{2n}}|$, where $j_1, \dots, j_{2n}$ is any permutation of $1, \dots, 2n$.
	Therefore,
	\begin{equation}
		\kappa_n \sim \frac{e^{cmr}}{N^{n-1}} \,,
		\label{eq:alpha0}
	\end{equation}
	where $c>0$ is an $\mathcal{O}(1)$ constant and $r$ is the characteristic separation between the points---the relative bias grows exponentially with $mr$ at $r\gtrsim 1/m$.
	Additional enhancements arise from the $(\Omega_\alpha)^{n-1}$ prefactor in Eq.~\eqref{eq:Gi2n}, as $\Omega_0 \sim \log(\Lambda/m)$ for $d=2$ and $\Omega_0 \sim (\Lambda/m)^{d-2}$ for $d\ge 3$.
	\item For $\alpha < 0$, the integrand in Eq.~\eqref{eq:Gi2n} is proportional to $1/(k^2+m^2)^{1+|\alpha|(n-1)}$, which can be written as derivatives of $1/(k^2+m^2)$ with respect to $m^2$.
	Taking $m^2$-derivatives does not change the exponential falloff of the integral, so Eq.~\eqref{eq:alpha0} still holds, with additional enhancements from $\Omega_\alpha\sim (\Lambda/m)^{d-2+2|\alpha|}$.
	\item For $\alpha > 0$, it is possible to choose the UV regulator $f_\Lambda$ such that the integral is exponentially suppressed at $r\gg 1/\Lambda$; e.g., $f_\Lambda(k^2) = e^{-k^2/2\Lambda^2}$ gives $G_i^{(2n)} \sim e^{-\Lambda^2 |x_\mathrm{in}-x_\mathrm{out}|^2/2}$.
	However, we do not expect the suppression to persist in the interacting theory, because integrating over internal coordinates of interaction vertices will pick up contributions from regions where $|x_\mathrm{in}-x_\mathrm{out}|\lesssim 1/\Lambda$.
\end{itemize}


\newpar
\paragraph{Noise}---\,
The finite-$N$ bias discussed above is systematic and can be removed by repeating the calculation at several values of $N$ and extrapolating to $N\to\infty$.
We now turn to the variance component of the error, which unlike the bias cannot be extrapolated away.

Generally, for any Monte Carlo approach that estimates correlation functions by averaging over $M$ samples of the random function $\phi$,
\begin{equation}
\hat{G}^{(2n)}(x_1, \dots, x_{2n}) \equiv \frac{1}{M} \sum_{m=1}^M \phi^{(m)}(x_1) \cdots \phi^{(m)}(x_{2n})\,,
\end{equation}
the variance of the estimator $\hat{G}^{(2n)}$ is given by
\begin{align}
&\mathrm{Var}\bigl[\hat{G}^{(2n)}(x_1, \dots, x_{2n})\bigr] \nonumber\\[2pt]
&= \frac{1}{M}\Bigl[\langle \phi(x_1)^2 \dots \phi(x_{2n})^2 \rangle - \langle\phi(x_1)\cdots\phi(x_{2n})\rangle^2\Bigr]\,.
\label{eq:variance}
\end{align}
The first term is a $4n$-point function with coincident points, which does not fall off exponentially and dominates the variance at large distances.
In NNFT with i.i.d.\ neurons, we can express this $4n$-point function in terms of single-neuron correlation functions similar to Eqs.~\eqref{eq:G4} and \eqref{eq:G6}; see Appendix~\ref{app:correlation-functions}.
In the $N\to\infty$ limit,
\begin{equation}
\mathrm{Var}\bigl[\hat{G}^{(2n)}(x_1, \dots, x_{2n})\bigr] \simeq \frac{1}{M} \prod_{j=1}^{2n} G^{(2)}_i(x_j,x_j) =\frac{(\Omega_0)^{2n}}{M} \,.
\label{eq:noisefloor}
\end{equation}
Eq.~\eqref{eq:noisefloor} defines an $\alpha$-independent noise floor, and implies a signal-to-noise ratio (SNR) that falls off exponentially with $mr$ at large distances:
\begin{equation}
\mathrm{SNR} \sim \sqrt{M}\, \frac{G^{(2n)}}{(\Omega_0)^n} \sim \sqrt{M}\, e^{-nmr}\,.
\label{eq:SNR}
\end{equation}
To achieve $\mathcal{O}(1)$ SNR, we therefore need $M \gtrsim e^{2nmr}$.
This is the same signal-to-noise problem that was first analyzed by Parisi and Lepage in the context of lattice field theory~\cite{Parisi:1983ae,Lepage:1989hd}.
The architecture-independent noise floor represents a fundamental cost of the NNFT approach.

On the other hand, at finite $N$, there are additional contributions to the variance, making Eq.~\eqref{eq:SNR} an upper bound on the achievable SNR.
We find that, similar to the bias, the size of finite-$N$ contributions relative to the noise floor is controlled by the ratios $\kappa_n$ defined in Eq.~\eqref{eq:kappa}, now evaluated at an even number of points that are pairwise coincident to yield
\begin{equation}
\kappa_n^\mathrm{noise} = \biggl(\frac{1}{2N}\biggr)^{\!n-1}\,\frac{(\Omega_\alpha)^{n-1}\,\Omega_{-\alpha(n-1)}}{(\Omega_0)^n}\,.
\label{eq:kappa2}
\end{equation}
From the definition of $\Omega_\alpha$ in Eq.~\eqref{eq:Omega}, it is straightforward to show that Eq.~\eqref{eq:kappa2} is minimized at $\alpha=0$ for all $n\ge 2$; see Appendix~\ref{app:kappa-noise} for details.
For $\alpha=0$, $\kappa_n^\mathrm{noise} = 1/(2N)^{n-1} \ll 1$ and we nearly saturate the bound in Eq.~\eqref{eq:SNR}; for either $\alpha<0$ or $\alpha>0$, it is possible for the finite-$N$ contributions to dominate over the noise floor due to enhancements by powers of $\Lambda/m$.

\begin{figure*}
\includegraphics[width=\linewidth]{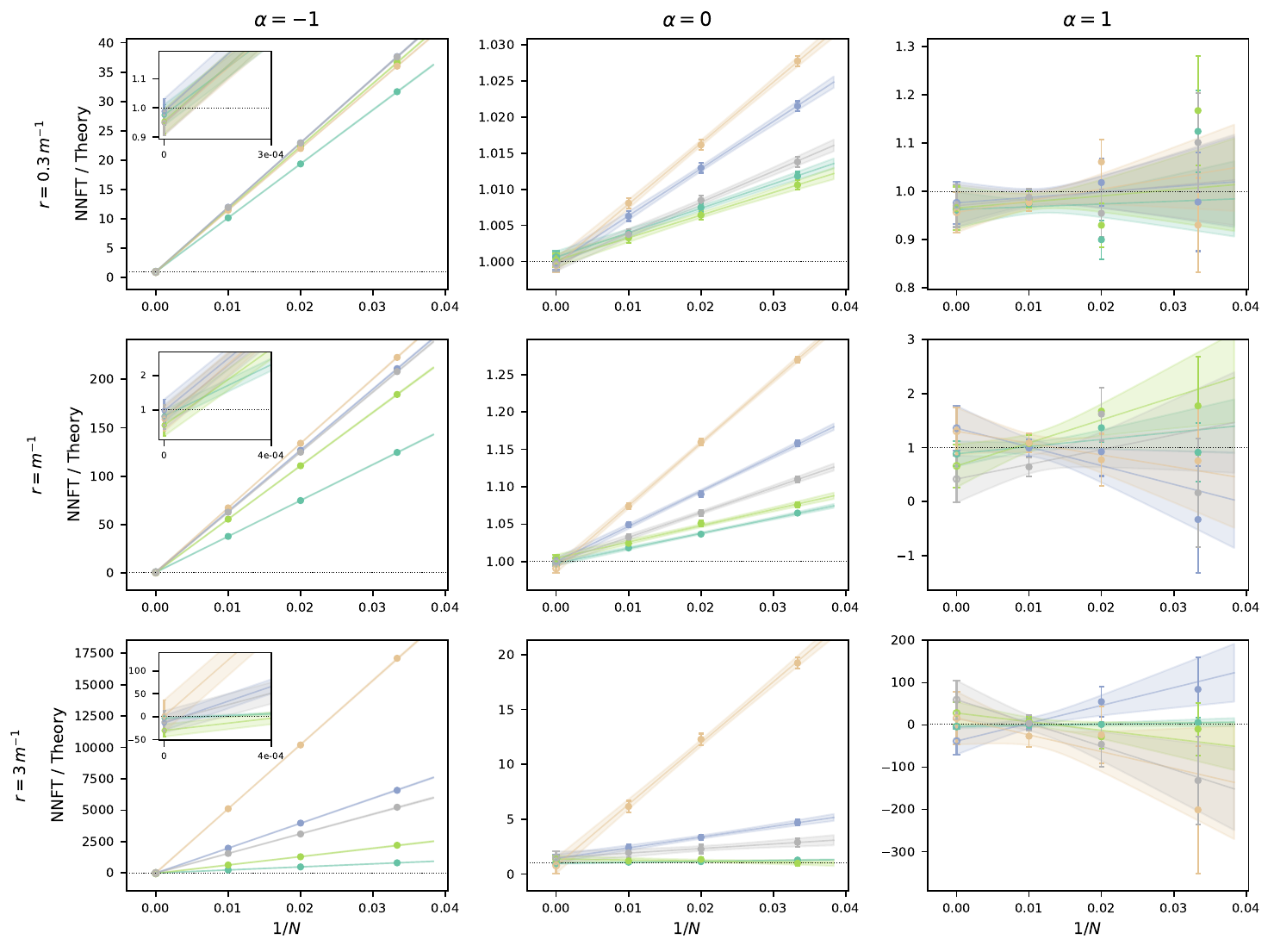}
\caption{Ratio of the NNFT four-point function to the free-theory prediction as a function of $1/N$, for 5 random configurations. Columns correspond to $\alpha = -1, 0, 1$; rows to characteristic separations $r = 0.3\,m^{-1}$, $m^{-1}$, $3\,m^{-1}$. Shaded bands indicate weighted linear fits and open circles mark the extrapolated intercepts at $1/N = 0$. Dotted line marks NNFT/Theory $= 1$. While previous work adopted $\alpha=-1$, we find that $\alpha=0$ is optimal in minimizing finite-$N$ errors.}
\label{fig:extrapolation}
\end{figure*}

\newpar
\paragraph{Numerical results}---\,
To test the analytic predictions above, we evaluate $G^{(4)}$ for 50 randomly generated four-point configurations in $d=2$ using an ensemble of $M=10^8$ Cos-Nets at each width $N = 30$, $50$, $100$.
We choose a Gaussian regulator, $f_\Lambda(k^2) = e^{-k^2/2\Lambda^2}$, with $\Lambda = 100\,m$.
For each configuration, we rescale the coordinates of the four points by a common factor to achieve three characteristic separations $r = 0.3\,m^{-1}$, $m^{-1}$, and $3\,m^{-1}$, where $r$ is defined as the average of the six pairwise distances between the four points.
Fig.~\ref{fig:extrapolation} shows the ratio of the NNFT result to the free-theory prediction as a function of $1/N$ for different values of $\alpha$ and $r$ for the first 5 configurations; the same qualitative behavior is observed for the other 45 configurations.
A weighted linear fit in $1/N$ is extrapolated to $1/N = 0$.

The contrast between different $\alpha$ is striking.
For $\alpha = 0$ (center column), the free-theory prediction is reproduced within much smaller uncertainties than the other two cases upon extrapolation to $1/N = 0$, confirming the optimality of this choice.
For $\alpha = -1$ (left column), the finite-$N$ ratios deviate from unity by orders of magnitude, reflecting $(\Lambda/m)$-enhanced bias relative to the $\alpha=0$ case.
For $\alpha = 1$ (right column), the ratios are consistent with unity at all $N$ within large error bars, confirming the suppressed bias but enhanced variance discussed above.
In all cases, we see a substantial increase in both finite-$N$ bias and variance with $r$, consistent with the analytically predicted exponential growth.
Across all 50 configurations and all $(\alpha, r)$ combinations, the extrapolated intercepts are consistent with unity within uncertainties, validating the $N$-extrapolation approach for removing bias.
The irreducible noise floor, on the other hand, presents an obstacle to achieving high precision at $r\gtrsim m^{-1}$.
We have also performed numerical experiments for $d=1, 3, 4$ and observed the same qualitative behavior as in $d=2$.


\newpar
\paragraph{Interacting theory}---\,
We can extend the bias and variance analysis to the interacting theory via Eq.~\eqref{eq:reweight}.
Consider $\phi^4$ theory as an example, with $S_\mathrm{int} = \frac{\lambda}{4!} \int d^dx\, \phi(x)^4$.
To analytically probe the $\alpha$-dependence, we focus on the perturbative regime and expand the right-hand side of Eq.~\eqref{eq:reweight} in $\lambda$.
The leading correction to the two-point function is given by
\begin{align}
	\bigl\langle \phi(x_1) \phi(x_2)\bigr\rangle\Bigr|_{\mathcal{O}(\lambda)} &= - \frac{\lambda}{4!}\, \int\! d^dx\, \Bigl[G^{(6)}(x_1, x_2, x, x, x, x) \nonumber\\[2pt]
	& - G^{(2)}(x_1,x_2) \,G^{(4)}(x, x, x, x)\Bigr]\,,
	\label{eq:phi4}
\end{align}
where $G^{(2)}$, $G^{(4)}$ and $G^{(6)}$ are the free-theory correlation functions.
Substituting in Eqs.~\eqref{eq:G4} and \eqref{eq:G6} and integrating over the vertex position $x$, we find that a subset of the finite-$N$ corrections have integrands independent of $x$, yielding contributions proportional to the spatial volume~\cite{Sen:2025vzl}.
At $\mathcal{O}(1/N)$, these are proportional to
\begin{equation}
	\Omega_\alpha\,\Omega_{-\alpha}(x_1-x_2) - \Omega_0\,\Omega_0(x_1-x_2)\,,
	\label{eq:phi4bias}
\end{equation}
where $\Omega_\alpha(r) \equiv \int \frac{d^dk}{(2\pi)^d} \frac{f_\Lambda(k^2)}{(k^2+m^2)^{\alpha+1}} e^{ik\cdot r}$ generalizes $\Omega_\alpha$ in Eq.~\eqref{eq:Omega} to nonzero $r$.
These and analogous contributions at $\mathcal{O}(1/N^2)$ vanish if and only if $\alpha=0$\,; see Appendix~\ref{app:phi4} for details.
This provides an independent argument for the optimality of $\alpha=0$, complementing the free-theory analysis.
Even for $\alpha=0$, the remaining finite-$N$ corrections include contributions that fall off more slowly than the field-theory prediction, and the relative bias grows exponentially with $mr$ as in the free theory.
We have verified that both the cancellation of IR divergences at $\alpha=0$ and the exponential growth of relative bias persist for the four-point function up to $\mathcal{O}(\lambda^2)$; a detailed analysis will be presented in a forthcoming paper.


\newpar
\paragraph{Summary and outlook}---\,
In this letter, we have identified a previously unexplored architectural freedom---parameterized by $\alpha$\,---in NNFT, a recently proposed approach for simulating field theories using neural network ensembles.
While all choices of $\alpha$ yield the same field theory in the limit of infinite network width $N\to\infty$, they give rise to drastically different finite-$N$ behaviors.
We have shown that $\alpha=0$, corresponding to sampling neuron momenta in direct proportion to the propagator while keeping the neuron amplitudes constant, is optimal for minimizing finite-$N$ errors in the evaluation of correlation functions.

Even for the optimal $\alpha=0$ architecture, relative errors from both bias and variance grow exponentially with distance $r$ at $r\gtrsim 1/m$\,---a qualitatively more severe limitation than the polynomial $\Lambda/m$ enhancement previously identified at $r\sim 1/m$.
The bias can be removed by extrapolating to $N\to\infty$, as we have demonstrated numerically.
The variance, on the other hand, sets an upper bound on the achievable SNR and imposes an exponential cost for measuring correlation functions at large distances, just as in lattice field theory.

Together, these results establish a practical strategy for turning NNFT from a proof-of-principle construction into a computational tool: adopt the optimal $\alpha=0$ architecture, extrapolate to $N\to\infty$ to remove bias, and maximize the ensemble size to suppress the noise floor.
That the signal-to-noise problem is shared with lattice field theory suggests that techniques developed in that context, such as variance reduction methods, may be applicable to NNFT as well.
The combination of these techniques with the continuum formulation intrinsic to NNFT may ultimately open a new frontier in nonperturbative field theory computations.


\begin{acknowledgments}
\newpar
\paragraph{Acknowledgments}---\,
Z.Z.\ is supported by the U.S.\ National Science Foundation under grant PHY-2412880.
This work has benefited from interactions with Claude Opus 4.5/4.6/4.7, which wrote the code for the numerical experiments under Z.Z.'s supervision and provided useful feedback on the manuscript; we thank Mehmet Demirtas and Ben Michel for facilitating initial access to Claude Max.
Simulation code, data and additional figures for all 50 four-point configurations are available at \url{https://github.com/zzkevin2019/nnft-alpha}.
\end{acknowledgments}

\bibliography{letter}

\begin{thebibliography}{31}%
\makeatletter
\providecommand \@ifxundefined [1]{%
 \@ifx{#1\undefined}
}%
\providecommand \@ifnum [1]{%
 \ifnum #1\expandafter \@firstoftwo
 \else \expandafter \@secondoftwo
 \fi
}%
\providecommand \@ifx [1]{%
 \ifx #1\expandafter \@firstoftwo
 \else \expandafter \@secondoftwo
 \fi
}%
\providecommand \natexlab [1]{#1}%
\providecommand \enquote  [1]{``#1''}%
\providecommand \bibnamefont  [1]{#1}%
\providecommand \bibfnamefont [1]{#1}%
\providecommand \citenamefont [1]{#1}%
\providecommand \href@noop [0]{\@secondoftwo}%
\providecommand \href [0]{\begingroup \@sanitize@url \@href}%
\providecommand \@href[1]{\@@startlink{#1}\@@href}%
\providecommand \@@href[1]{\endgroup#1\@@endlink}%
\providecommand \@sanitize@url [0]{\catcode `\\12\catcode `\$12\catcode `\&12\catcode `\#12\catcode `\^12\catcode `\_12\catcode `\%12\relax}%
\providecommand \@@startlink[1]{}%
\providecommand \@@endlink[0]{}%
\providecommand \url  [0]{\begingroup\@sanitize@url \@url }%
\providecommand \@url [1]{\endgroup\@href {#1}{\urlprefix }}%
\providecommand \urlprefix  [0]{URL }%
\providecommand \Eprint [0]{\href }%
\providecommand \doibase [0]{https://doi.org/}%
\providecommand \selectlanguage [0]{\@gobble}%
\providecommand \bibinfo  [0]{\@secondoftwo}%
\providecommand \bibfield  [0]{\@secondoftwo}%
\providecommand \translation [1]{[#1]}%
\providecommand \BibitemOpen [0]{}%
\providecommand \bibitemStop [0]{}%
\providecommand \bibitemNoStop [0]{.\EOS\space}%
\providecommand \EOS [0]{\spacefactor3000\relax}%
\providecommand \BibitemShut  [1]{\csname bibitem#1\endcsname}%
\let\auto@bib@innerbib\@empty
\bibitem [{\citenamefont {Halverson}\ \emph {et~al.}(2021)\citenamefont {Halverson}, \citenamefont {Maiti},\ and\ \citenamefont {Stoner}}]{Halverson:2020trp}%
  \BibitemOpen
  \bibfield  {author} {\bibinfo {author} {\bibfnamefont {J.}~\bibnamefont {Halverson}}, \bibinfo {author} {\bibfnamefont {A.}~\bibnamefont {Maiti}},\ and\ \bibinfo {author} {\bibfnamefont {K.}~\bibnamefont {Stoner}},\ }\bibfield  {title} {\bibinfo {title} {{Neural Networks and Quantum Field Theory}},\ }\href {https://doi.org/10.1088/2632-2153/abeca3} {\bibfield  {journal} {\bibinfo  {journal} {Mach. Learn. Sci. Tech.}\ }\textbf {\bibinfo {volume} {2}},\ \bibinfo {pages} {035002} (\bibinfo {year} {2021})},\ \Eprint {https://arxiv.org/abs/2008.08601} {arXiv:2008.08601 [cs.LG]} \BibitemShut {NoStop}%
\bibitem [{\citenamefont {Halverson}(2021)}]{Halverson:2021aot}%
  \BibitemOpen
  \bibfield  {author} {\bibinfo {author} {\bibfnamefont {J.}~\bibnamefont {Halverson}},\ }\bibfield  {title} {\bibinfo {title} {{Building Quantum Field Theories Out of Neurons}}\ }(\bibinfo {year} {2021})\ \Eprint {https://arxiv.org/abs/2112.04527} {arXiv:2112.04527 [hep-th]} \BibitemShut {NoStop}%
\bibitem [{\citenamefont {Demirtas}\ \emph {et~al.}(2024)\citenamefont {Demirtas}, \citenamefont {Halverson}, \citenamefont {Maiti}, \citenamefont {Schwartz},\ and\ \citenamefont {Stoner}}]{Demirtas:2023fir}%
  \BibitemOpen
  \bibfield  {author} {\bibinfo {author} {\bibfnamefont {M.}~\bibnamefont {Demirtas}}, \bibinfo {author} {\bibfnamefont {J.}~\bibnamefont {Halverson}}, \bibinfo {author} {\bibfnamefont {A.}~\bibnamefont {Maiti}}, \bibinfo {author} {\bibfnamefont {M.~D.}\ \bibnamefont {Schwartz}},\ and\ \bibinfo {author} {\bibfnamefont {K.}~\bibnamefont {Stoner}},\ }\bibfield  {title} {\bibinfo {title} {{Neural network field theories: non-Gaussianity, actions, and locality}},\ }\href {https://doi.org/10.1088/2632-2153/ad17d3} {\bibfield  {journal} {\bibinfo  {journal} {Mach. Learn. Sci. Tech.}\ }\textbf {\bibinfo {volume} {5}},\ \bibinfo {pages} {015002} (\bibinfo {year} {2024})},\ \Eprint {https://arxiv.org/abs/2307.03223} {arXiv:2307.03223 [hep-th]} \BibitemShut {NoStop}%
\bibitem [{\citenamefont {Halverson}\ \emph {et~al.}(2025)\citenamefont {Halverson}, \citenamefont {Naskar},\ and\ \citenamefont {Tian}}]{Halverson:2024axc}%
  \BibitemOpen
  \bibfield  {author} {\bibinfo {author} {\bibfnamefont {J.}~\bibnamefont {Halverson}}, \bibinfo {author} {\bibfnamefont {J.}~\bibnamefont {Naskar}},\ and\ \bibinfo {author} {\bibfnamefont {J.}~\bibnamefont {Tian}},\ }\bibfield  {title} {\bibinfo {title} {{Conformal fields from neural networks}},\ }\href {https://doi.org/10.1007/JHEP10(2025)039} {\bibfield  {journal} {\bibinfo  {journal} {JHEP}\ }\textbf {\bibinfo {volume} {10}},\ \bibinfo {pages} {039}},\ \Eprint {https://arxiv.org/abs/2409.12222} {arXiv:2409.12222 [hep-th]} \BibitemShut {NoStop}%
\bibitem [{\citenamefont {Capuozzo}\ \emph {et~al.}(2025)\citenamefont {Capuozzo}, \citenamefont {Robinson},\ and\ \citenamefont {Suzzoni}}]{Capuozzo:2025ozt}%
  \BibitemOpen
  \bibfield  {author} {\bibinfo {author} {\bibfnamefont {P.}~\bibnamefont {Capuozzo}}, \bibinfo {author} {\bibfnamefont {B.}~\bibnamefont {Robinson}},\ and\ \bibinfo {author} {\bibfnamefont {B.}~\bibnamefont {Suzzoni}},\ }\bibfield  {title} {\bibinfo {title} {{Conformal Defects in Neural Network Field Theories}}\ }(\bibinfo {year} {2025})\ \Eprint {https://arxiv.org/abs/2512.07946} {arXiv:2512.07946 [hep-th]} \BibitemShut {NoStop}%
\bibitem [{\citenamefont {Robinson}(2025)}]{Robinson:2025ybg}%
  \BibitemOpen
  \bibfield  {author} {\bibinfo {author} {\bibfnamefont {B.}~\bibnamefont {Robinson}},\ }\bibfield  {title} {\bibinfo {title} {{Virasoro Symmetry in Neural Network Field Theories}}\ }(\bibinfo {year} {2025})\ \Eprint {https://arxiv.org/abs/2512.24420} {arXiv:2512.24420 [hep-th]} \BibitemShut {NoStop}%
\bibitem [{\citenamefont {Huang}\ and\ \citenamefont {Zhou}(2026)}]{Huang:2025ipy}%
  \BibitemOpen
  \bibfield  {author} {\bibinfo {author} {\bibfnamefont {G.}~\bibnamefont {Huang}}\ and\ \bibinfo {author} {\bibfnamefont {K.}~\bibnamefont {Zhou}},\ }\bibfield  {title} {\bibinfo {title} {{The neural networks with tensor weights and emergent fermionic Wick rules in the large-width limit}},\ }\href {https://doi.org/10.1016/j.physletb.2025.140146} {\bibfield  {journal} {\bibinfo  {journal} {Phys. Lett. B}\ }\textbf {\bibinfo {volume} {873}},\ \bibinfo {pages} {140146} (\bibinfo {year} {2026})},\ \Eprint {https://arxiv.org/abs/2507.05303} {arXiv:2507.05303 [hep-th]} \BibitemShut {NoStop}%
\bibitem [{\citenamefont {Frank}\ \emph {et~al.}(2025)\citenamefont {Frank}, \citenamefont {Halverson}, \citenamefont {Maiti},\ and\ \citenamefont {Ruehle}}]{Frank:2025zuk}%
  \BibitemOpen
  \bibfield  {author} {\bibinfo {author} {\bibfnamefont {S.}~\bibnamefont {Frank}}, \bibinfo {author} {\bibfnamefont {J.}~\bibnamefont {Halverson}}, \bibinfo {author} {\bibfnamefont {A.}~\bibnamefont {Maiti}},\ and\ \bibinfo {author} {\bibfnamefont {F.}~\bibnamefont {Ruehle}},\ }\bibfield  {title} {\bibinfo {title} {{Fermions and Supersymmetry in Neural Network Field Theories}}\ }(\bibinfo {year} {2025})\ \Eprint {https://arxiv.org/abs/2511.16741} {arXiv:2511.16741 [hep-th]} \BibitemShut {NoStop}%
\bibitem [{\citenamefont {Frank}\ and\ \citenamefont {Halverson}(2026)}]{Frank:2026bui}%
  \BibitemOpen
  \bibfield  {author} {\bibinfo {author} {\bibfnamefont {S.}~\bibnamefont {Frank}}\ and\ \bibinfo {author} {\bibfnamefont {J.}~\bibnamefont {Halverson}},\ }\bibfield  {title} {\bibinfo {title} {{String Theory from Infinite Width Neural Networks}}\ }(\bibinfo {year} {2026})\ \Eprint {https://arxiv.org/abs/2601.06249} {arXiv:2601.06249 [hep-th]} \BibitemShut {NoStop}%
\bibitem [{\citenamefont {Ageev}\ and\ \citenamefont {Ageeva}(2026{\natexlab{a}})}]{Ageev:2026ofv}%
  \BibitemOpen
  \bibfield  {author} {\bibinfo {author} {\bibfnamefont {D.~S.}\ \bibnamefont {Ageev}}\ and\ \bibinfo {author} {\bibfnamefont {Y.~A.}\ \bibnamefont {Ageeva}},\ }\bibfield  {title} {\bibinfo {title} {{Excited String States and D-branes from Infinite Width Neural Networks}}\ }(\bibinfo {year} {2026})\ \Eprint {https://arxiv.org/abs/2602.10214} {arXiv:2602.10214 [hep-th]} \BibitemShut {NoStop}%
\bibitem [{\citenamefont {Erbin}\ \emph {et~al.}(2022{\natexlab{a}})\citenamefont {Erbin}, \citenamefont {Lahoche},\ and\ \citenamefont {Samary}}]{Erbin:2021kqf}%
  \BibitemOpen
  \bibfield  {author} {\bibinfo {author} {\bibfnamefont {H.}~\bibnamefont {Erbin}}, \bibinfo {author} {\bibfnamefont {V.}~\bibnamefont {Lahoche}},\ and\ \bibinfo {author} {\bibfnamefont {D.~O.}\ \bibnamefont {Samary}},\ }\bibfield  {title} {\bibinfo {title} {{Non-perturbative renormalization for the neural network-QFT correspondence}},\ }\href {https://doi.org/10.1088/2632-2153/ac4f69} {\bibfield  {journal} {\bibinfo  {journal} {Mach. Learn. Sci. Tech.}\ }\textbf {\bibinfo {volume} {3}},\ \bibinfo {pages} {015027} (\bibinfo {year} {2022}{\natexlab{a}})},\ \Eprint {https://arxiv.org/abs/2108.01403} {arXiv:2108.01403 [hep-th]} \BibitemShut {NoStop}%
\bibitem [{\citenamefont {Erbin}\ \emph {et~al.}(2022{\natexlab{b}})\citenamefont {Erbin}, \citenamefont {Lahoche},\ and\ \citenamefont {Samary}}]{Erbin:2022lls}%
  \BibitemOpen
  \bibfield  {author} {\bibinfo {author} {\bibfnamefont {H.}~\bibnamefont {Erbin}}, \bibinfo {author} {\bibfnamefont {V.}~\bibnamefont {Lahoche}},\ and\ \bibinfo {author} {\bibfnamefont {D.~O.}\ \bibnamefont {Samary}},\ }\bibfield  {title} {\bibinfo {title} {{Renormalization in the neural network-quantum field theory correspondence}}\ }(\bibinfo {year} {2022})\ \Eprint {https://arxiv.org/abs/2212.11811} {arXiv:2212.11811 [hep-th]} \BibitemShut {NoStop}%
\bibitem [{\citenamefont {Howard}\ \emph {et~al.}(2025)\citenamefont {Howard}, \citenamefont {Klinger}, \citenamefont {Maiti},\ and\ \citenamefont {Stapleton}}]{Howard:2024kfd}%
  \BibitemOpen
  \bibfield  {author} {\bibinfo {author} {\bibfnamefont {J.~N.}\ \bibnamefont {Howard}}, \bibinfo {author} {\bibfnamefont {M.~S.}\ \bibnamefont {Klinger}}, \bibinfo {author} {\bibfnamefont {A.}~\bibnamefont {Maiti}},\ and\ \bibinfo {author} {\bibfnamefont {A.~G.}\ \bibnamefont {Stapleton}},\ }\bibfield  {title} {\bibinfo {title} {{Bayesian RG flow in neural network field theories}},\ }\href {https://doi.org/10.21468/SciPostPhysCore.8.1.027} {\bibfield  {journal} {\bibinfo  {journal} {SciPost Phys. Core}\ }\textbf {\bibinfo {volume} {8}},\ \bibinfo {pages} {027} (\bibinfo {year} {2025})},\ \Eprint {https://arxiv.org/abs/2405.17538} {arXiv:2405.17538 [hep-th]} \BibitemShut {NoStop}%
\bibitem [{\citenamefont {Halverson}(2024)}]{Halverson:2024hax}%
  \BibitemOpen
  \bibfield  {author} {\bibinfo {author} {\bibfnamefont {J.}~\bibnamefont {Halverson}},\ }\bibfield  {title} {\bibinfo {title} {{TASI Lectures on Physics for Machine Learning}}\ }(\bibinfo {year} {2024})\ \Eprint {https://arxiv.org/abs/2408.00082} {arXiv:2408.00082 [hep-th]} \BibitemShut {NoStop}%
\bibitem [{\citenamefont {Ferko}\ and\ \citenamefont {Halverson}(2026)}]{Ferko:2025ogz}%
  \BibitemOpen
  \bibfield  {author} {\bibinfo {author} {\bibfnamefont {C.}~\bibnamefont {Ferko}}\ and\ \bibinfo {author} {\bibfnamefont {J.}~\bibnamefont {Halverson}},\ }\bibfield  {title} {\bibinfo {title} {{Quantum mechanics and neural networks}},\ }\href {https://doi.org/10.1088/2632-2153/ae3105} {\bibfield  {journal} {\bibinfo  {journal} {Mach. Learn. Sci. Tech.}\ }\textbf {\bibinfo {volume} {7}},\ \bibinfo {pages} {015002} (\bibinfo {year} {2026})},\ \Eprint {https://arxiv.org/abs/2504.05462} {arXiv:2504.05462 [hep-th]} \BibitemShut {NoStop}%
\bibitem [{\citenamefont {Ferko}\ \emph {et~al.}(2026{\natexlab{a}})\citenamefont {Ferko}, \citenamefont {Halverson},\ and\ \citenamefont {Mutchler}}]{Ferko:2026axm}%
  \BibitemOpen
  \bibfield  {author} {\bibinfo {author} {\bibfnamefont {C.}~\bibnamefont {Ferko}}, \bibinfo {author} {\bibfnamefont {J.}~\bibnamefont {Halverson}},\ and\ \bibinfo {author} {\bibfnamefont {A.}~\bibnamefont {Mutchler}},\ }\bibfield  {title} {\bibinfo {title} {{Universality of Neural Network Field Theory}}\ }(\bibinfo {year} {2026})\ \Eprint {https://arxiv.org/abs/2601.14453} {arXiv:2601.14453 [hep-th]} \BibitemShut {NoStop}%
\bibitem [{\citenamefont {Ageev}\ and\ \citenamefont {Ageeva}(2026{\natexlab{b}})}]{Ageev:2026qyh}%
  \BibitemOpen
  \bibfield  {author} {\bibinfo {author} {\bibfnamefont {D.~S.}\ \bibnamefont {Ageev}}\ and\ \bibinfo {author} {\bibfnamefont {Y.~A.}\ \bibnamefont {Ageeva}},\ }\bibfield  {title} {\bibinfo {title} {{Neural Network Quantum Field Theory from Transformer Architectures}}\ }(\bibinfo {year} {2026})\ \Eprint {https://arxiv.org/abs/2602.10209} {arXiv:2602.10209 [cs.LG]} \BibitemShut {NoStop}%
\bibitem [{\citenamefont {Ferko}\ \emph {et~al.}(2026{\natexlab{b}})\citenamefont {Ferko}, \citenamefont {Halverson}, \citenamefont {Jejjala},\ and\ \citenamefont {Robinson}}]{Ferko:2026ken}%
  \BibitemOpen
  \bibfield  {author} {\bibinfo {author} {\bibfnamefont {C.}~\bibnamefont {Ferko}}, \bibinfo {author} {\bibfnamefont {J.}~\bibnamefont {Halverson}}, \bibinfo {author} {\bibfnamefont {V.}~\bibnamefont {Jejjala}},\ and\ \bibinfo {author} {\bibfnamefont {B.}~\bibnamefont {Robinson}},\ }\bibfield  {title} {\bibinfo {title} {{Topological Effects in Neural Network Field Theory}}\ }(\bibinfo {year} {2026})\ \Eprint {https://arxiv.org/abs/2604.02313} {arXiv:2604.02313 [hep-th]} \BibitemShut {NoStop}%
\bibitem [{\citenamefont {Maiti}\ \emph {et~al.}(2021)\citenamefont {Maiti}, \citenamefont {Stoner},\ and\ \citenamefont {Halverson}}]{Maiti:2021fpy}%
  \BibitemOpen
  \bibfield  {author} {\bibinfo {author} {\bibfnamefont {A.}~\bibnamefont {Maiti}}, \bibinfo {author} {\bibfnamefont {K.}~\bibnamefont {Stoner}},\ and\ \bibinfo {author} {\bibfnamefont {J.}~\bibnamefont {Halverson}},\ }\bibfield  {title} {\bibinfo {title} {{Symmetry-via-Duality: Invariant Neural Network Densities from Parameter-Space Correlators}}\ }(\bibinfo {year} {2021})\ \Eprint {https://arxiv.org/abs/2106.00694} {arXiv:2106.00694 [cs.LG]} \BibitemShut {NoStop}%
\bibitem [{\citenamefont {Neal}(1996)}]{Neal1996}%
  \BibitemOpen
  \bibfield  {author} {\bibinfo {author} {\bibfnamefont {R.~M.}\ \bibnamefont {Neal}},\ }\href@noop {} {\emph {\bibinfo {title} {Bayesian Learning for Neural Networks}}},\ \bibinfo {series} {Lecture Notes in Statistics}, Vol.\ \bibinfo {volume} {118}\ (\bibinfo  {publisher} {Springer},\ \bibinfo {year} {1996})\BibitemShut {NoStop}%
\bibitem [{\citenamefont {Williams}(1996)}]{Williams1996}%
  \BibitemOpen
  \bibfield  {author} {\bibinfo {author} {\bibfnamefont {C.~K.~I.}\ \bibnamefont {Williams}},\ }\bibfield  {title} {\bibinfo {title} {Computing with infinite networks},\ }in\ \href {https://proceedings.neurips.cc/paper/1996/file/ae5e3ce40e0404a45ecacaaf05e5f735-Paper.pdf} {\emph {\bibinfo {booktitle} {Advances in Neural Information Processing Systems}}},\ Vol.~\bibinfo {volume} {9},\ \bibinfo {editor} {edited by\ \bibinfo {editor} {\bibfnamefont {M.}~\bibnamefont {Mozer}}, \bibinfo {editor} {\bibfnamefont {M.}~\bibnamefont {Jordan}},\ and\ \bibinfo {editor} {\bibfnamefont {T.}~\bibnamefont {Petsche}}}\ (\bibinfo  {publisher} {MIT Press},\ \bibinfo {year} {1996})\BibitemShut {NoStop}%
\bibitem [{\citenamefont {Lee}\ \emph {et~al.}(2017)\citenamefont {Lee}, \citenamefont {Bahri}, \citenamefont {Novak}, \citenamefont {Schoenholz}, \citenamefont {Pennington},\ and\ \citenamefont {Sohl-Dickstein}}]{Lee:2017qzq}%
  \BibitemOpen
  \bibfield  {author} {\bibinfo {author} {\bibfnamefont {J.}~\bibnamefont {Lee}}, \bibinfo {author} {\bibfnamefont {Y.}~\bibnamefont {Bahri}}, \bibinfo {author} {\bibfnamefont {R.}~\bibnamefont {Novak}}, \bibinfo {author} {\bibfnamefont {S.~S.}\ \bibnamefont {Schoenholz}}, \bibinfo {author} {\bibfnamefont {J.}~\bibnamefont {Pennington}},\ and\ \bibinfo {author} {\bibfnamefont {J.}~\bibnamefont {Sohl-Dickstein}},\ }\bibfield  {title} {\bibinfo {title} {{Deep Neural Networks as Gaussian Processes}}\ }(\bibinfo {year} {2017})\ \Eprint {https://arxiv.org/abs/1711.00165} {arXiv:1711.00165 [stat.ML]} \BibitemShut {NoStop}%
\bibitem [{\citenamefont {Matthews}\ \emph {et~al.}(2018)\citenamefont {Matthews}, \citenamefont {Rowland}, \citenamefont {Hron}, \citenamefont {Turner},\ and\ \citenamefont {Ghahramani}}]{Matthews2018}%
  \BibitemOpen
  \bibfield  {author} {\bibinfo {author} {\bibfnamefont {A.~G. d.~G.}\ \bibnamefont {Matthews}}, \bibinfo {author} {\bibfnamefont {M.}~\bibnamefont {Rowland}}, \bibinfo {author} {\bibfnamefont {J.}~\bibnamefont {Hron}}, \bibinfo {author} {\bibfnamefont {R.~E.}\ \bibnamefont {Turner}},\ and\ \bibinfo {author} {\bibfnamefont {Z.}~\bibnamefont {Ghahramani}},\ }\bibfield  {title} {\bibinfo {title} {Gaussian process behaviour in wide deep neural networks}\ }(\bibinfo {year} {2018})\ \Eprint {https://arxiv.org/abs/1804.11271} {arXiv:1804.11271 [stat.ML]} \BibitemShut {NoStop}%
\bibitem [{\citenamefont {Yang}(2019)}]{Yang2019}%
  \BibitemOpen
  \bibfield  {author} {\bibinfo {author} {\bibfnamefont {G.}~\bibnamefont {Yang}},\ }\bibfield  {title} {\bibinfo {title} {Tensor programs i: Wide feedforward or recurrent neural networks of any architecture are gaussian processes}\ }(\bibinfo {year} {2019})\ \Eprint {https://arxiv.org/abs/1910.12478} {arXiv:1910.12478 [cs.NE]} \BibitemShut {NoStop}%
\bibitem [{\citenamefont {Hanin}(2021)}]{Hanin2021}%
  \BibitemOpen
  \bibfield  {author} {\bibinfo {author} {\bibfnamefont {B.}~\bibnamefont {Hanin}},\ }\bibfield  {title} {\bibinfo {title} {Random neural networks in the infinite width limit as gaussian processes}\ }(\bibinfo {year} {2021})\ \Eprint {https://arxiv.org/abs/2107.01562} {arXiv:2107.01562 [math.PR]} \BibitemShut {NoStop}%
\bibitem [{\citenamefont {Sen}\ and\ \citenamefont {Vaidya}(2025)}]{Sen:2025vzl}%
  \BibitemOpen
  \bibfield  {author} {\bibinfo {author} {\bibfnamefont {S.}~\bibnamefont {Sen}}\ and\ \bibinfo {author} {\bibfnamefont {V.}~\bibnamefont {Vaidya}},\ }\bibfield  {title} {\bibinfo {title} {{Viability of perturbative expansion for quantum field theories on neurons}}\ }(\bibinfo {year} {2025})\ \Eprint {https://arxiv.org/abs/2508.03810} {arXiv:2508.03810 [hep-th]} \BibitemShut {NoStop}%
\bibitem [{\citenamefont {Rahimi}\ and\ \citenamefont {Recht}(2007)}]{Rahimi2007}%
  \BibitemOpen
  \bibfield  {author} {\bibinfo {author} {\bibfnamefont {A.}~\bibnamefont {Rahimi}}\ and\ \bibinfo {author} {\bibfnamefont {B.}~\bibnamefont {Recht}},\ }\bibfield  {title} {\bibinfo {title} {Random features for large-scale kernel machines},\ }in\ \href {https://proceedings.neurips.cc/paper/2007/hash/013a006f03dbc5392effeb8f18fda755-Abstract.html} {\emph {\bibinfo {booktitle} {Advances in Neural Information Processing Systems}}},\ Vol.~\bibinfo {volume} {20},\ \bibinfo {editor} {edited by\ \bibinfo {editor} {\bibfnamefont {J.}~\bibnamefont {Platt}}, \bibinfo {editor} {\bibfnamefont {D.}~\bibnamefont {Koller}}, \bibinfo {editor} {\bibfnamefont {Y.}~\bibnamefont {Singer}},\ and\ \bibinfo {editor} {\bibfnamefont {S.}~\bibnamefont {Roweis}}}\ (\bibinfo  {publisher} {Curran Associates, Inc.},\ \bibinfo {year} {2007})\ pp.\ \bibinfo {pages} {1177--1184}\BibitemShut {NoStop}%
\bibitem [{Note1()}]{Note1}%
  \BibitemOpen
  \bibinfo {note} {This freedom was noted in passing in Ref.~\cite {Frank:2026bui} in the context of the 2d free boson.}\BibitemShut {Stop}%
\bibitem [{Note2()}]{Note2}%
  \BibitemOpen
  \bibinfo {note} {Sampling $|a_i|$ from a Gaussian as in Refs.~\cite {Halverson:2021aot,Demirtas:2023fir} instead gives $\langle |a_i|^{2n}\rangle = (2n-1)!!\protect \,\langle |a_i|^2\rangle ^n$, which amplifies the bias and variance by numerical prefactors but does not change the parametric dependence on $\alpha $.}\BibitemShut {Stop}%
\bibitem [{\citenamefont {Parisi}(1984)}]{Parisi:1983ae}%
  \BibitemOpen
  \bibfield  {author} {\bibinfo {author} {\bibfnamefont {G.}~\bibnamefont {Parisi}},\ }\bibfield  {title} {\bibinfo {title} {{The Strategy for Computing the Hadronic Mass Spectrum}},\ }\href {https://doi.org/10.1016/0370-1573(84)90081-4} {\bibfield  {journal} {\bibinfo  {journal} {Phys. Rept.}\ }\textbf {\bibinfo {volume} {103}},\ \bibinfo {pages} {203} (\bibinfo {year} {1984})}\BibitemShut {NoStop}%
\bibitem [{\citenamefont {Lepage}(1989)}]{Lepage:1989hd}%
  \BibitemOpen
  \bibfield  {author} {\bibinfo {author} {\bibfnamefont {G.~P.}\ \bibnamefont {Lepage}},\ }\bibfield  {title} {\bibinfo {title} {{The Analysis of Algorithms for Lattice Field Theory}},\ }in\ \href@noop {} {\emph {\bibinfo {booktitle} {{Theoretical Advanced Study Institute in Elementary Particle Physics}}}}\ (\bibinfo {year} {1989})\BibitemShut {NoStop}%
\end{thebibliography}%

\section*{End Matter}

\appsection{Correlation functions for i.i.d.\ neurons}
\label{app:correlation-functions}
To derive the four-point function in Eq.~\eqref{eq:G4}, we start from $\phi(x) = \frac{1}{\sqrt{N}}\sum_i \phi_i(x)$ and write
\begin{equation}
G^{(4)}(x_1, \dots, x_4) = \frac{1}{N^2} \sum_{ijkl} \bigl\langle \phi_i(x_1) \phi_j(x_2) \phi_k(x_3) \phi_l(x_4)\bigr\rangle\,.
\end{equation}
There are two types of contributions.
First, when $i,j,k,l$ are pairwise equal, we obtain the product of two single-neuron two-point functions.
There are $N(N-1)$ such terms in the sum for each of the three distinct pairings, and they give identical contributions by the i.i.d.\ assumption, yielding the first term in Eq.~\eqref{eq:G4}.
Second, when $i,j,k,l$ are all equal, we obtain the single-neuron four-point function.
There are $N$ such terms in the sum, yielding the second term in Eq.~\eqref{eq:G4}.
The six-point function in Eq.~\eqref{eq:G6} can be derived similarly by considering the various ways of grouping the six $\phi_i$'s and counting combinatorial factors.
The same procedure applies to higher-point functions, as discussed previously in Ref.~\cite{Sen:2025vzl}.

For the coincident-point $4n$-point function that appears in the variance Eq.~\eqref{eq:variance}, the leading contribution comes from terms where the pair of $\phi_i$'s at the same $x$ point have the same neuron index.
In the $N\to\infty$ limit, this gives Eq.~\eqref{eq:noisefloor}; all other pairings of neuron indices give products of single-neuron two-point functions evaluated at non-coincident points which are smaller than the coincident-point two-point functions.
Similarly, finite-$N$ corrections are dominated by terms where the single-neuron correlation functions, $G^{(2n)}_i$ given in Eq.~\eqref{eq:Gi2n}, are evaluated at points that are pairwise coincident; in this case there exists a balanced partition such that $x_\mathrm{in} = x_\mathrm{out}$, and we arrive at the $\kappa_n^\mathrm{noise}$ formula in Eq.~\eqref{eq:kappa2} upon substituting in the definition of $\Omega_\alpha$ from Eq.~\eqref{eq:Omega}.

\newpar
\appsection{Minimization of $\kappa_n^\mathrm{noise}$}
\label{app:kappa-noise}
To minimize $\kappa_n^\mathrm{noise}$ given by Eq.~\eqref{eq:kappa2} with respect to $\alpha$, we denote the $\alpha$-derivatives of $\Omega_\alpha$ by $\Omega'_\alpha$, $\Omega''_\alpha$, etc.
Differentiating $(\Omega_\alpha)^{n-1}\,\Omega_{-\alpha(n-1)}$ with respect to $\alpha$, we find:
\begin{align}
&\frac{d}{d\alpha}\Bigl[(\Omega_\alpha)^{n-1}\,\Omega_{-\alpha(n-1)}\Bigr] \nonumber\\[2pt]
&= (n-1) \Bigl[(\Omega_\alpha)^{n-2}\,\Omega_{-\alpha(n-1)}\,\Omega'_\alpha -(\Omega_\alpha)^{n-1}\,\Omega'_{-\alpha(n-1)}\Bigr]\,,
\end{align}
which vanishes at $\alpha=0$.
Taking the second derivative, we find:
\begin{align}
&\frac{d^2}{d\alpha^2}\Bigl[(\Omega_\alpha)^{n-1}\,\Omega_{-\alpha(n-1)}\Bigr]\Bigr|_{\alpha=0} \nonumber\\[2pt]
&= n(n-1)\,(\Omega_0)^{n-2}\,\Bigl[\Omega_0\,\Omega''_0 - (\Omega'_0)^2\Bigr]\,.
\end{align}
By the Cauchy–Schwarz inequality,
\begin{align}
&\Omega_0\,\Omega''_0 - (\Omega'_0)^2 \nonumber\\[2pt]
&= \int \frac{d^dk'}{(2\pi)^d} \frac{f_\Lambda(k'^2)}{(k'^2+m^2)} \int \frac{d^dk}{(2\pi)^d} \frac{f_\Lambda(k^2)}{(k^2+m^2)}\,\log^2(k^2+m^2) \nonumber\\
&\quad - \biggl[\int \frac{d^dk}{(2\pi)^d} \frac{f_\Lambda(k^2)}{(k^2+m^2)}\,\log(k^2+m^2)\biggr]^2 \ge 0\,,
\end{align}
which shows that $\alpha=0$ is a minimum of $\kappa_n^\mathrm{noise}$ for all $n\ge 2$.

\newpar
\appsection{Two-point function in $\phi^4$ theory}
\label{app:phi4}
We can evaluate the $\mathcal{O}(\lambda)$ contribution to the two-point function in $\phi^4$ theory, given in Eq.~\eqref{eq:phi4}, by substituting in Eqs.~\eqref{eq:G4} and \eqref{eq:G6} together with $G^{(2)}(x_1,x_2) = G_i^{(2)}(x_1,x_2)$, and using Eq.~\eqref{eq:Gi2n} for the single-neuron correlation functions.
To simplify notation, we note that, with $\Omega_\alpha(r)$ introduced below Eq.~\eqref{eq:phi4bias}, Eq.~\eqref{eq:Gi2n} can be written as
\begin{equation}
G_i^{(2n)}\!(x_1, \dots, x_{2n}) = \frac12 \biggl(\frac{\Omega_\alpha}{2}\biggr)^{\!\!n-1}\! \sum_\mathrm{P} \Omega_{-\alpha(n-1)}(x_\mathrm{in}-x_\mathrm{out}) \,.
\label{eq:Gi2n2}
\end{equation}
This equation is valid for all $n\ge 1$; for $n=1$ it reduces to $G_i^{(2)}(x_1,x_2) = \Omega_0(x_1-x_2)$.
Noting also that $\Omega_\alpha(0) = \Omega_\alpha$, we find:
\begin{widetext}
\begin{align}
\bigl\langle \phi(x_1) \phi(x_2)\bigr\rangle\Bigr|_{\mathcal{O}(\lambda)} &= -\lambda \int\! d^dx\, \biggl\{ \biggl(1-\frac{1}{N}\biggr) \biggl(1-\frac{2}{N}\biggr)\,\frac12\, \Omega_0\,\Omega_0(x-x_1)\,\Omega_0(x-x_2) \nonumber\\[2pt]
&\;\; + \frac{1}{N} \biggl(1-\frac{1}{N}\biggr) \biggl[ \,\frac14\,\Omega_\alpha \Bigl( \Omega_0(x-x_1)\,\Omega_{-\alpha}(x-x_2) + \Omega_0(x-x_2)\,\Omega_{-\alpha}(x-x_1) \Bigr) \nonumber\\
&\qquad\qquad\qquad\qquad +\hl{\frac18\,\Omega_0\,\Omega_\alpha} \Bigl( \hl{2\,\Omega_{-\alpha}(x_1- x_2)} +\Omega_{-\alpha}(2x-x_1-x_2) \Bigr) \,\hl{-\,\frac14\,(\Omega_0)^2\,\Omega_0(x_1-x_2)} \biggr] \nonumber\\
&\;\; +\frac{1}{N^2}\,\frac{1}{4!} \biggl[ \hl{\frac12\,(\Omega_\alpha)^2} \Bigl( \hl{3\,\Omega_{-2\alpha}(x_1-x_2)} +2\,\Omega_{-2\alpha}(2x-x_1-x_2) \Bigr) \,\hl{-\,\frac32\,\Omega_\alpha\,\Omega_{-\alpha}\,\Omega_0(x_1-x_2)} \biggr] \biggr\}\,.
\label{eq:phi4full}
\end{align}
\end{widetext}
The first line is the field-theory prediction rescaled by a factor of $1+\mathcal{O}(1/N)$, while the remaining lines are additional finite-$N$ corrections.
The IR-divergent terms, highlighted in \hl{red}, are those where the dependence on the integration variable $x$ cancels out between $x_\mathrm{in}$ and $x_\mathrm{out}$.
The presence of these terms at $\mathcal{O}(1/N)$ was noted in Ref.~\cite{Sen:2025vzl} for $\alpha=-1$.
From Eq.~\eqref{eq:phi4full}, we see that the IR-divergent terms cancel if and only if $\alpha=0$, in which case we have:
\begin{align}
&\bigl\langle \phi(x_1) \phi(x_2)\bigr\rangle\Bigr|_{\mathcal{O}(\lambda)} \nonumber\\
&= -\lambda \int\! d^dx\, \biggl\{ \biggl(1-\frac{1}{N}\biggr)^2\,\frac12\, \Omega_0\,\Omega_0(x-x_1)\,\Omega_0(x-x_2) \nonumber\\[2pt]
&\qquad + \frac{1}{N} \biggl(1-\frac{2}{3N}\biggr)\, \frac18\,(\Omega_0)^2\,\Omega_0(2x-x_1-x_2) \biggr\}\,.
\end{align}
The last term has an integrand that falls off more slowly than the field-theory result at large distances because $\Omega_0(2x-x_1-x_2) \sim e^{-m|2x-x_1-x_2|}$, and $|2x-x_1-x_2| \le |x-x_1| + |x-x_2|$ by the triangle inequality, just like in our free-theory analysis.
In fact, shifting the integration variable from $x$ to $x-\frac{x_1+x_2}{2}$, we see that this term yields a constant that is independent of $x_1$, $x_2$, and thus constitutes a bias that grows exponentially with $mr$ relative to the field-theory prediction at large distances.

\end{document}